\documentclass[conference]{IEEEtran}
\IEEEoverridecommandlockouts

\usepackage{cite}
\usepackage{amsmath,amssymb,amsfonts}
\usepackage{algorithmic}
\usepackage{graphicx}
\usepackage{textcomp}
\usepackage{xcolor}
\usepackage{booktabs}
\usepackage{cleveref}
\usepackage{xspace}
\usepackage{tikz}
\usepackage{multirow}
\usepackage{enumitem}
\usepackage{balance}
\usepackage{wrapfig}
\usepackage{needspace}
\usepackage{tcolorbox}
\usepackage{xurl}
\usepackage{multicol}
\usepackage{textpos}
\usepackage{pifont}

\graphicspath{{figures/}}

\def\BibTeX{{\rm B\kern-.05em{\sc i\kern-.025em b}\kern-.08em
    T\kern-.1667em\lower.7ex\hbox{E}\kern-.125emX}}
\definecolor{aliceblue}{rgb}{0.94, 0.97, 1.0}
\tcbset{width=1\columnwidth,boxrule=1pt,colback=aliceblue,arc=1pt,left=2pt,right=2pt,boxsep=0.5pt}

\newcommand\revhl[1]{{#1}}

\newcommand{\checkvalue}[1]{{#1}}

\newcommand{\papername}{SlipStream\xspace}

\newcommand{\parhead}[1]{\noindent\textbf{#1}}

\Crefname{figure}{Fig.}{Figs.}
\crefname{figure}{Fig.}{Figs.}
\crefformat{section}{#2\S#1#3}
\crefformat{subsection}{#2\S#1#3}
\crefformat{subsubsection}{#2\S#1#3}

\makeatletter
\def\@IEEENORMtitlevspace{0pt}
\def\@IEEEMINtitlevspace{0pt}
\def\ps@firstpage{%
  \ps@plain
  \def\@oddhead{\hfil{\normalfont
    To appear in 2027 IEEE International Symposium on High-Performance Computer Architecture (HPCA)}\hfil}%
  \let\@evenhead\@oddhead
}
\makeatother

\begin{document}

\title{\papername: Enabling HBF for GPU Systems by Mitigating the Latency and SRAM Overheads}

\author{%
\IEEEauthorblockN{Anish Saxena\textsuperscript{*}\thanks{\textsuperscript{*}Anish can be reached at \texttt{asaxena317@gatech.edu}.}}
\IEEEauthorblockA{\textit{Georgia Tech}\\
Atlanta, GA, USA}
\and
\IEEEauthorblockN{Hritvik Taneja}
\IEEEauthorblockA{\textit{Georgia Tech}\\
Atlanta, GA, USA}
\and
\IEEEauthorblockN{Christos Kozyrakis}
\IEEEauthorblockA{\textit{NVIDIA Research \& Stanford University }\\
Santa Clara, CA, USA}
\and
\IEEEauthorblockN{Aamer Jaleel}
\IEEEauthorblockA{\textit{NVIDIA Research}\\
Westford, MA, USA}
}

\maketitle
\thispagestyle{firstpage}
\pagestyle{plain}

\setcounter{page}{1}
\begin{abstract}

Large language model (LLM) inference increasingly requires memory
capacity for larger models and batch sizes. Emerging High-Bandwidth Flash (HBF) technology
replaces HBM's DRAM dies with NAND flash, providing up to 16$\times$
the capacity at comparable read bandwidth. However, HBF incurs a nearly 20$\,\mu$s read latency, 20$\times$ that of HBM, 
causing on-demand HBF accesses to slow inference by up to
12.8$\times$. Conventional cache-line prefetchers require $20\times$ more in-flight
tracking state to cover HBF latency, while tensor-streaming designs such
as $H^3$ follow virtual-address rather than kernel-demand order and
therefore need hundreds of megabytes of buffering to perform similarly to an idealized HBF system with HBM-like latency.

We propose \papername, a software--hardware co-designed prefetcher
that turns prefetching into a demand-paced producer--consumer staging
pipeline. 
\papername's software component profiles the repeatable per-CTA weight-page demand order of data-independent LLM kernels and inserts a \textit{staging warp} to replay this order with non-blocking \texttt{STAGE} instructions. 
\papername's hardware component provides architectural support for the staging warp, treating it as the producer and the work warps as consumers. Our key
insight is to retire each \texttt{STAGE} tracking entry only when demand first accesses the page, rather than when the page returns from HBF, so demand itself paces staging, bounding how far staging can lead consumption. \papername tracks 4\,KB pages instead of cache lines, and allocates
staging buffer slots only when pages return, thereby decoupling its size from outstanding traffic. Across seven dense models spanning 16--800\,GB,
\papername with a 40\,MB staging buffer 
achieves 92.4\% of ideal-HBF performance on average, whereas $H^3$ retains only 43.1\% performance of ideal-HBF (i.e., 2.3$\times$ slowdown), even with a 320\,MB staging buffer. 
\papername's buffer sizing is independent of HBF latency and workload characteristics, and \papername with a small 16MB staging buffer---$20\times$ lower SRAM than $H^3$---performs within 4\% of \papername with 40MB buffer.
\revhl{Finally, \papername supports Mixture-of-Experts models and reaches 91.2\% of ideal-HBF with a 40\,MB buffer.}

\end{abstract}

\section{Introduction}

Large language model (LLM) inference increasingly demands memory capacity without a corresponding demand for additional computation. Model weights alone can occupy hundreds of gigabytes; for example, Llama~3.1 405B requires approximately 405GB even with one-byte weights~\cite{grattafiori2024llama}. Long-context and high-concurrency serving further increase the memory footprint through key-value (KV) caches~\cite{vllm,zheng2024sglang,gemini1.5, guo2025deepseek}. Systems today obtain the required capacity by distributing inference across many GPUs~\cite{sglang_pd_disagg, shoeybi2019megatron, pope2023efficiently, gb200_whitepaper}. During bandwidth-bound decoding, however, much of the compute provided by these additional GPUs remains underutilized~\cite{patel2024splitwise, zhong2024distserve}.

Memory vendors are developing new technologies that scale both memory bandwidth and capacity, such as High-Bandwidth Flash (HBF). HBF replaces the DRAM dies in an HBM stack with NAND flash to provide up to 16$\times$ the capacity of HBM at comparable bandwidth~\cite{hbf,h3_llm}. This additional capacity comes with four important limitations: high read latency, low write bandwidth, high power consumption, and limited write endurance. The write-bandwidth and endurance constraints can be avoided by using HBF for read-only data.
LLM weights are therefore particularly well suited for placement in HBF in proposed HBM-HBF systems~\cite{h3_llm, hsu2026haven}.
Doing so also frees up HBM capacity to store more KV state and improve decoding throughput, which can improve energy efficiency (e.g., lower EDP or higher throughput per watt). 
Thus, HBF offers the potential to provide substantially greater GPU memory capacity and improve LLM inference.

\begin{figure*}[!t]
  \centering
  \includegraphics[width=\textwidth]{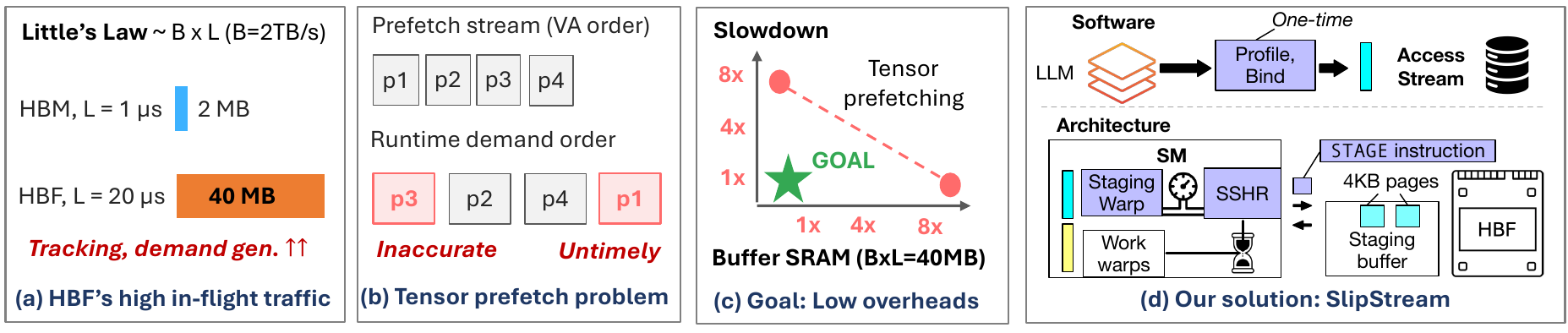}
  \caption{(a) Increasing load-to-use latency from 1\,$\mu$s with HBM to 20\,$\mu$s with HBF increases the required outstanding data by 20$\times$.  (b) Prefetching tensor pages by virtual address does not reproduce the order in which each kernel requests them, resulting in inaccurate and untimely prefetching.  (c) Goal: minimize slowdown, tracking SRAM, and buffer SRAM.  (d) \papername uses software guidance and hardware flow control to bound tracking and buffer SRAM.}
  \label{fig:visual_abstract}
\end{figure*}

Read latency, however, remains a fundamental obstacle to using HBF in GPUs. An HBF round-trip access can incur 20$\times$ the latency of HBM, 20$\mu$s compared to about $1\mu$s on current GPUs~\cite{h3_llm}. By Little's Law, sustaining a fixed bandwidth across a 20$\times$ longer latency requires 20$\times$ more traffic in flight, which current GPUs cannot maintain.
In our simulations, replacing HBM with iso-bandwidth HBF on A100 GPU slows inference of Llama-8B by 7.0$\times$ and Llama-70B by 11.1$\times$. 
Increasing in-flight traffic requires a corresponding increase in miss-tracking state (\Cref{fig:visual_abstract}(a)). Typical miss-status holding registers (MSHRs) perform associative lookup, making a proportional increase expensive in both area and complexity.

Moreover, increasing tracking state alone does not solve the HBF latency problem: the GPU must also generate enough independent memory requests to keep the enlarged memory pipeline full. Existing SMs do not expose this degree of demand-side memory-level parallelism (e.g., due to limited shared memory), and high-end GPUs already approach reticle-scale die limits, leaving little opportunity to increase SMs and MSHRs merely to accommodate flash latency.

A long-standing approach to tolerating memory latency is to prefetch data before the corresponding demand requests arrive. Unfortunately, conventional GPU prefetchers issue requests at cache-line granularity and do not scale to the substantially long HBF latency. Extending these prefetchers to HBF would still require proportionally larger MSHRs and request queues for tracking, along with more buffering for returned data. Recent HBF proposals instead prefetch entire tensors by streaming their pages sequentially into a staging buffer~\cite{h3_llm}.
This approach requires either a buffer large enough to retain the tensor or an accurate prefetch stream that matches the page-demand order.
In practice, software tiling causes GPU kernels to consume tensor pages in an order that can differ substantially from their in-memory layout order (e.g., virtual-address order), as \Cref{fig:visual_abstract}(b) shows.
With a smaller buffer than tensor's footprint, demand might outrun the prefetch stream, resulting in exposed HBF latency. In our studies, such a design with 40\,MB buffer leaves Llama-70B at 10.2$\times$ the execution time of an ideal-HBF system (same latency as HBM), and even a  320\,MB buffer still incurs 2.3$\times$ slowdown.

Our goal is to exploit the capacity and bandwidth of HBF for LLM
inference at near-HBM performance with low tracking and SRAM
overhead (\Cref{fig:visual_abstract}(c)) while minimizing changes to GPU programming model.
To this end, we propose
\emph{\papername}, a software--hardware co-designed prefetcher in
which a software-profiled access stream runs ahead of
demand warps, while demand-based feedback bounds the
gap between producer and consumer. Software determines \emph{what} pages
to stage and in \emph{what order}; hardware determines \emph{how far ahead}
staging may run and manages the returned data. 
We develop \papername, shown in \Cref{fig:visual_abstract}(d), based on three insights.

First, software constructs an accurate staging stream. The staging order must follow the
kernel's page-demand order, which can differ substantially from the
tensor's in-memory layout. We find that this demand order is repeatable
for data-independent LLM kernels: they request weight pages in an
identical per-CTA order across inputs and decode steps. Software can
therefore profile this order once and cache it as a compact access stream. At
runtime, a \textit{staging warp} added to each CTA replays the stream as
non-blocking \texttt{STAGE} instructions retrieving 4\,KB pages from HBF.
Because the staging warp resides in the same CTA as the
consumer work warps, it is mapped to the same SM. 
Its
nonblocking, page-granular \texttt{STAGE} instructions allow
it to advance rapidly through the exact sequence that the work
warps subsequently demand.

Second, hardware keeps staging of access stream timely by pacing it with
consumer progress. A page staged too late
exposes HBF latency, whereas one staged too early occupies buffer space
before its consumer arrives and may displace useful data; either outcome
can stall an SM for microseconds. Conventional prefetchers estimate a
prefetch distance and tolerate occasional mistakes, but an uncovered HBF
access can expose up to 20\,$\mu$s of latency. \papername instead treats
the staging warp as a producer and the work warps as consumers. Each
\texttt{STAGE} allocates a page-granular \textit{Stage Status Holding Register} (SSHR) entry, and a full
SSHR backpressures only the staging warp. Unlike a conventional Miss Status Holding Register (MSHR),
which releases its entry when the requested data returns, \papername
retains each SSHR entry until the page's first demand access. 
When a staged page serves its first demand, \papername
releases the SSHR entry, enabling the staging
warp to make progress. SSHR occupancy therefore measures the hardware-bounded
producer--consumer gap and ties staging progress to
consumer progress.

Third, a page-granular software--hardware interface makes this
long-latency staging pipeline inexpensive.
At the evaluated 2\,TB/s design point, a 20\,$\mu$s lookahead spans
approximately 40\,MB of data---about 9,800 4\,KB pages, or more than
300,000 128-byte cache lines. Tracking this window at cache-line
granularity with MSHRs would be impractical.
Fortunately, as HBF transfers 4\,KB pages, one \texttt{STAGE}
instruction's tracking entry can represent an entire page,
reducing required tracking entries by 32$\times$.
Moreover, we only use a 4KB slot in the staging SRAM buffer when the page actually arrives from HBF, and in-flight pages consume only compact tracking metadata. 
We make buffer eviction policy consumption-aware to shrink its size  relative to volume of in-flight data.

Across seven dense LLMs running on Ampere, Hopper, and Blackwell GPUs, replaying \papername's profiled streams adds 1--3\% runtime overhead on GPU silicon. In cycle-level simulation with these models (16--800GB in size) and a 40\,MB staging
buffer, \papername retains 92.4\% performance of ideal-HBF, whereas the
$H^3$ tensor prefetcher retains just 43.1\% even with a 320\,MB
buffer. 
\revhl{\papername also supports Mixture-of-Experts models. On two Qwen3 MoEs (60--470GB), it reaches 91.2\% of ideal-HBF.}
Finally, demand-paced staging and consumption-aware buffering further allow \papername to work with a small 16MB buffer at just 12.6\% slowdown.

\noindent Overall, this paper makes the following contributions:
\begin{itemize}[leftmargin=*,itemsep=0pt,topsep=2pt]
  \item We find that iso-bandwidth HBF requires 20$\times$ more
  outstanding data than HBM, beyond what GPUs can sustain.

  \item We develop \papername's software that profiles
  LLMs to produce access streams for runtime replay via staging warps.

  \item We devise an efficient caching scheme so streams remain
  reusable across different LLMs and GPU architectures.

  \item We architect \papername's hardware, in which first-demand feedback
  paces the staging warp's lead over demand.

  \item We leverage demand-paced staging and buffer eviction
  to further reduce buffer size with negligible runtime overheads.

\end{itemize}

\section{Background and Motivation}
\label{sec:background}

\subsection{LLM Inference and the GPU Memory Wall}

LLM inference demands more memory capacity and bandwidth to
increase model size and batch more requests. During inference, compute-bound \emph{prefill} phase
processes an input sequence in parallel, while the memory-bound
\emph{decode} phase generates one token at a time; modern deployments disaggregate the two phases~\cite{zhong2024distserve, patel2024splitwise, sglang_pd_disagg, nvl72}.
Each decode iteration of a dense
model reads all the model weights irrespective of the batch size \revhl{(\cref{sec:moe_support} discusses Mixture-of-Experts~\cite{shazeer2017moe})}. So, the
weight footprint and read traffic grow with
model size~\cite{grattafiori2024llama}, and require high-capacity, high-bandwidth memory.

\subsection{Emerging High-Bandwidth Flash (HBF) Technology}
\label{sec:hbf_substrate}

Recently, memory vendors have proposed High-Bandwidth Flash (HBF), where DRAM dies in an
HBM stack are replaced with NAND flash while retaining an
HBM-like base die~\cite{hbf,h3_llm}. This organization provides $16{\times}$ the capacity of HBM at comparable aggregate read bandwidth.
Table~\ref{tab:hbm_hbf} summarizes HBF (and HBM) properties in a GPU system.
However, the NAND substrate introduces four key challenges.

\parhead{\#1 Write endurance.} HBF's limited write endurance precludes storing
frequently updated data: NAND cells endure only $10^4$--$10^5$
program/erase cycles~\cite{luo2018nand}. Thus HBF would have 100--1000$\times$
lower write bandwidth than HBM (\Cref{tab:hbm_hbf}).
LLM weights are written once during model loading and read at
every decode step, so we use HBF exclusively for weights.

\parhead{\#2 Power consumption.} HBF can consume 4$\times$ more power than iso-bandwidth HBM~\cite{h3_llm}, but memory is a small fraction of a GPU's kW-scale power budget~\cite{blackwell}. \revhl{Moving weights to HBF frees HBM capacity for KV state and can significantly improve inference throughput at a lower EDP (\cref{sec:serving}).}

\parhead{\#3 Read granularity.} HBF returns a 4\,KB page per read. The SM should read the entire page shortly after it arrives to avoid wasting bandwidth.
SMs operate on compute tiles, so a tile-major layout,
where a page corresponds to a single tile, would ensure full page use,
and is already used by some LLM kernel
libraries~\cite{cute_tiled,tiling_cutlass,frantar2025marlin}. 
We assume \emph{consume-once} behavior: each page read is fully consumed shortly after arrival.

\parhead{\#4 Read latency.} HBF's NAND sensing, controller, and
error-correction overheads can yield read latency of up to $20\mu$s (SM load-to-use
latency), 20$\times$ that of HBM~\cite{cheong2018flash,h3_llm}, making it a fundamental
obstacle to using HBF for GPU systems.

\begin{table}[ht]
  \centering
  \caption{Memory device properties.}
  \label{tab:hbm_hbf}
  \begin{footnotesize}
  \begin{tabular}{@{}lcc@{}}
    \toprule
    \textbf{Property} & \textbf{HBM} & \textbf{HBF} \\
    \midrule
    Capacity per stack & 16GB & 256GB \\
    Aggregate read bandwidth & 2\,TB/s & 2\,TB/s \\
    SM load-to-use latency & ${\sim}$1\,$\mu$s & ${\sim}$20\,$\mu$s \\
    Read granularity & 32B line & 4\,KB page \\
    Outstanding data$^\dagger$ & $2\,MB$ & $40\,MB$ \\
    Power consumption & ${\sim}$40W & ${\sim}$160W \\
    Aggregate write bandwidth & 2\,TB/s & ${\sim}$1\,GB/s sustained$^\ddagger$ \\
    Write endurance & $\geq10^{15}$ accesses & $10^4$--$10^5$ P/E cycles \\
    \bottomrule
  \end{tabular}
  \end{footnotesize}
  \parbox{0.97\columnwidth}{\scriptsize $^\dagger$Little's Law (  \Cref{sec:hbf_latency}).  $^\ddagger$Assuming three-year service life and continuous writes.}
\end{table}

\subsection{Problem: Long HBF Read Latency}
\label{sec:hbf_latency}

\begin{figure*}[htb]
  \centering
  \includegraphics[width=\textwidth]{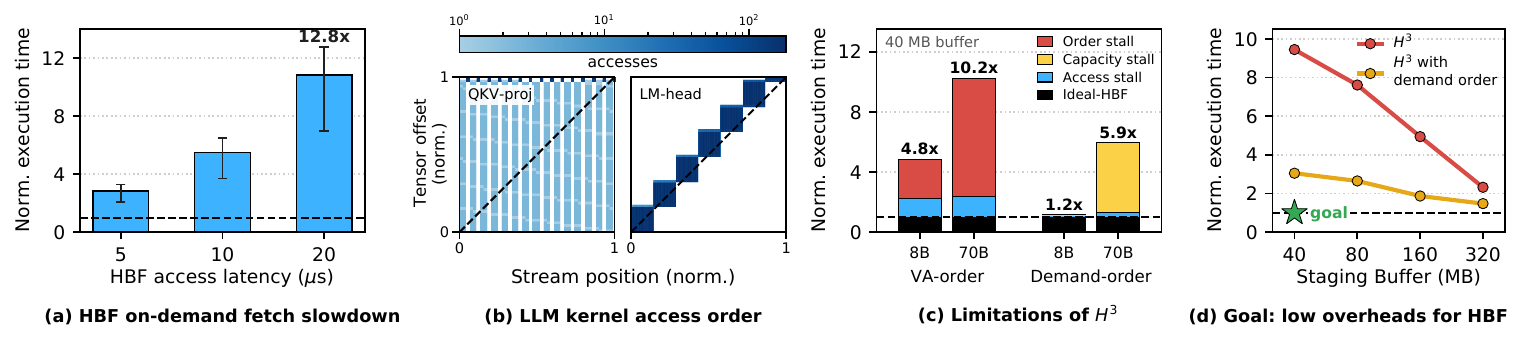}
\caption{ (a) \revhl{Execution time normalized to ideal-HBF} of LLM on-demand fetch from HBF for different HBF access latencies. 20$\mu$s access incurs 10.8x average slowdown, and up to 12.8x (seven LLMs, \Cref{sec:methodology}). 
 (b) Profiled accesses of two LLM kernels. LLM kernel accesses do not follow tensor's VA-order (x=y line). 
 (c) Performance of $H^3$ for Llama-8B and Llama-70B, when prefetcher streams in VA-order (high slowdown due to order stalls) and in demand-order (provided by \papername in \Cref{sec:design}, capacity stalls dominate).
 (d) Seven-model performance vs buffer SRAM for $H^3$. Our goal is
\revhl{ideal-HBF} parity within a small buffer.
}
  \label{fig:sec2_motivation}
\end{figure*}

By Little's Law, sustaining bandwidth $B$ across latency $L$ requires
$B \times L$ bytes in flight, so HBF requires 20$\times$ more outstanding
data than HBM---40\,MB total (\Cref{tab:hbm_hbf}). Today's GPUs are
not provisioned for this scale: MSHRs are sized for HBM's
$\approx$1\,$\mu$s load-to-use latency~\cite{a100, h100, blackwell}, and even the aggregate shared
memory (18\,MB/29\,MB on an A100/H100) falls short. Our
simulations of an A100-class GPU show that serving weights from HBF slows
LLM inference by 10.8$\times$ (geometric mean), and by up to
12.8$\times$, across seven LLMs, relative to an ideal-HBF system 
with same read latency as HBM. The slowdown grows with
latency across our 5--20\,$\mu$s sweep (\Cref{fig:sec2_motivation}(a)). 
Scaling request-side
structures (e.g., MSHRs) alone is insufficient---more MSHRs need more
demand-generating SMs to fill them, and modern GPUs are already
reticle-limited~\cite{a100}.
\revhl{In our simulations, 20$\times$ more MSHRs and queues improve
on-demand HBF to only 14.6\% of ideal-HBF performance, due to insufficient demand-side memory level parallelism. A larger L2 cache also fails
to help, because reuse distance of weights (across iterations) far exceeds cache capacity.
}
Using bulk-copy engines such as Tensor Memory Accelerator~\cite{h100} still requires reserving more shared memory than what current GPUs provision.
Hiding HBF latency therefore requires a
more efficient solution.

\subsection{Limitations of Existing Prefetching Techniques for HBF}
\label{sec:h3_limitations}

\subsubsection{Prefetching cachelines}
GPUs prefetch cache lines either transparently in hardware
~\cite{lee_mta,koo_cta_aware,snake} or explicitly through software
prefetch instructions~\cite{nvidia_ptx_isa}. 
However, line-granular prefetching inherits the
bookkeeping of the demand path, so covering a 20$\times$ longer access
requires 20$\times$ the tracking state (\Cref{sec:hbf_latency}). Moreover,
one tardy or inaccurate prefetch stalls the SM for microseconds, incurring drastic slowdown.
\revhl{\Cref{sec:related_work} provides quantitative comparison for various prior techniques.}

\subsubsection{Prefetching tensors}
\label{sec:tensor_prefetch}

A recent work, $H^3$, proposes prefetching at tensor-granularity, because ``LLM inference has
a deterministic and sequential access pattern''~\cite{h3_llm}. A software
hint names a tensor, and $H^3$ streams its pages into
a latency-hiding \textit{staging buffer}. \revhl{Unfortunately, $H^3$
does not describe its architecture, so we construct a capable and optimistic
design with lightweight flow control and sufficient MSHRs. It prefetches a page only when a buffer slot
is free (paces aggressiveness) and evicts only accessed pages (avoids thrashing)}\footnote{$H^3$ advocates double-buffering, which
permits cheaper single-ported SRAM at twice the capacity for similar runtime
overhead.}. Tensor prefetching hides HBF latency if the
prefetch stream matches the kernel's access order and the buffer retains
every fetched page until demand arrives. 
We find the first
condition fails outright and second requires impractical SRAM.

\parhead{Challenge \#1: matching kernel's access order.}
  $H^3$ prefetches pages in tensor's virtual-address (VA) order.
  However, LLM kernels partition each tensor among its
  concurrent CTAs and swizzle the mapping for locality~\cite{osama2023stream, frantar2025marlin}.
  \Cref{fig:sec2_motivation}(b) shows the demand order of two Llama-8B
  kernels: QKV-proj touches VAs across the entire tensor at nearly every stream
  position, LM-head walks blocks of rows, and neither kernel follows the VA
  diagonal.
  Thus, demand often reaches a page before the stream does, \revhl{so the load
  fetches the page itself} and waits the full 20\,$\mu$s; we call this an
  \emph{order stall}.
  With 40\,MB buffer, order stalls leave Llama-8B at
  4.8$\times$ and Llama-70B at 10.2$\times$ slowdown
  (\Cref{fig:sec2_motivation}(c)-left). An oracle stream that fetches pages in
  the demand order eliminates order stalls and recovers Llama-8B to
  \revhl{1.2}$\times$ \revhl{due to \textit{access stalls}, in which
  demand waits for pages that the stream has fetched but that are still in
  flight (mostly cold start).} Llama-70B, however,
  remains at \revhl{5.9}$\times$. 

  \parhead{Challenge \#2: retaining pages in the buffer.}
  Even with kernel's access order, prefetched pages must remain in the
  staging buffer until demand consumes them. Since the staging buffer is allocated
  on arrival from HBF and retained until consumption by demands, unconsumed pages
  can fill the staging buffer and prevent subsequent prefetched
  pages from being installed.
  When kernels exhibit cross-SM load imbalance \revhl{(e.g., the gate and up kernels
  of Llama-70B launch 224 equal-work CTAs on 108 SMs, so eight SMs run three CTAs
  and access 1.5$\times$ more pages than the other 100),} pages awaiting lagging SMs
  can occupy the majority of the shared buffer, leaving insufficient
  space for pages needed by faster SMs. The prefetcher then stalls
  until buffer slots become free, and the faster SMs eventually demand
  pages whose prefetch requests have not yet been issued, causing \textit{capacity stalls} (e.g., \revhl{5.9}$\times$ loss for Llama-70B in \Cref{fig:sec2_motivation}(c)).
  \revhl{This imbalance comes from grid quantization,
  so CTA-to-SM scheduling cannot remove it. It appears in the gate and up
  kernels of three of our seven models (with 55--60\% of their weight
  traffic).}
  Thus, the staging buffer requires \emph{demand-proportional
  allocation}: an SM should acquire additional capacity only as its
  demand consumes previously staged pages. Without such flow control,
  tensor prefetching must overprovision the buffer to absorb cross-SM
  consumption skew. Even with a 320\,MB buffer, the execution time is $1.48\times$ (\Cref{fig:sec2_motivation}(d)), and up to $4.7\times$.

\begin{tcolorbox}
\label{sec:motiv_goal}
\noindent{\bf Key Takeaway and Goal:} A practical HBF prefetcher must supply
the demand order of each kernel and decouple buffer capacity from the
tensor footprint. Our goal is to serve HBF accesses at near-HBM performance with low
storage overhead and minimal architectural and programming changes.
\end{tcolorbox}

\section{Experimental Methodology}
\label{sec:methodology}

\begin{table*}[!htb]
  \centering
  \caption{Workload characterization (per-layer, batch size 64, FP16).
  CTA counts include split-K slices (grid $z$). Time is in $\mu$s.}
  \label{tab:method_workloads}
  \begin{scriptsize}
  \setlength{\tabcolsep}{0.9pt}
  \renewcommand{\arraystretch}{1.05}
  \begin{tabular*}{\textwidth}{@{\extracolsep{\fill}}lrrcrrcrrcrrcrrcrrc@{}}
    \toprule
    & \multicolumn{3}{c}{\textbf{QKV-Proj}} &
      \multicolumn{3}{c}{\textbf{O-Proj}} &
      \multicolumn{3}{c}{\textbf{MLP-Gate}} &
      \multicolumn{3}{c}{\textbf{MLP-Up}} &
      \multicolumn{3}{c}{\textbf{MLP-Down}} &
      \multicolumn{3}{c}{\textbf{Total}} \\
    \cmidrule(lr){2-4}\cmidrule(lr){5-7}\cmidrule(lr){8-10}
    \cmidrule(lr){11-13}\cmidrule(lr){14-16}\cmidrule(l){17-19}
    \textbf{Model}
      & \textbf{Time} & \textbf{\shortstack{Wt.\\pages}} & \textbf{CTAs}
      & \textbf{Time} & \textbf{\shortstack{Wt.\\pages}} & \textbf{CTAs}
      & \textbf{Time} & \textbf{\shortstack{Wt.\\pages}} & \textbf{CTAs}
      & \textbf{Time} & \textbf{\shortstack{Wt.\\pages}} & \textbf{CTAs}
      & \textbf{Time} & \textbf{\shortstack{Wt.\\pages}} & \textbf{CTAs}
      & \textbf{Time} & \textbf{\shortstack{Wt.\\pages}} & \textbf{CTAs} \\
    \midrule
    Llama-8B~\cite{grattafiori2024llama}
      &  50.3 &  12K & 256 &  34.7 &   8K & 64 &   91.5 &  29K & 56 &   89.8 &  29K & 56 &   86.7 &  29K & 96 &  353.0 &  107K & 528 \\
    Mistral-24B~\cite{mistral_24b}
      &  60.2 &  15K & 256 &  41.0 &  10K & 80 &  219.5 &  82K & 256 &  218.9 &  82K & 256 &  234.2 &  82K & 100 &  773.8 &  271K & 948 \\
    Qwen2.5-32B~\cite{yang2024qwen2technicalreport}
      &  73.1 &  18K & 200 &  37.2 &  13K & 80 &  160.5 &  69K & 108 &  163.2 &  69K & 108 &  193.0 &  69K & 100 &  627.0 &  238K & 596 \\
    DeepSeek-67B~\cite{bi2024deepseek}
      & 121.1 &  41K & 272 & 117.3 &  33K & 64 &  242.4 &  88K & 86 &  245.6 &  88K & 86 &  238.1 &  88K & 96 &  964.5 &  338K & 604 \\
    Llama-70B~\cite{grattafiori2024llama}
      & 121.0 &  41K & 272 & 117.3 &  33K & 64 &  420.5 & 115K & 224 &  424.0 & 115K & 224 &  308.2 & 115K & 96 & 1391.0 &  418K & 880 \\
    Qwen2.5-72B~\cite{yang2024qwen2technicalreport}
      & 122.3 &  41K & 224 & 118.3 &  33K & 64 &  419.0 & 118K & 231 &  418.2 & 118K & 231 &  310.1 & 118K & 96 & 1387.9 &  429K & 846 \\
    Llama-405B~\cite{grattafiori2024llama}
      & 424.2 & 148K & 272 & 378.9 & 131K & 64 & 1001.5 & 426K & 416 & 1003.1 & 426K & 416 & 1226.3 & 426K & 64 & 4034.0 & 1557K & 1232 \\
    \bottomrule
  \end{tabular*}
  \end{scriptsize}
\end{table*}

\subsection{System organization}
\label{sec:sys_org}

\begin{figure}[htb]
  \centering
  \includegraphics[width=\columnwidth]{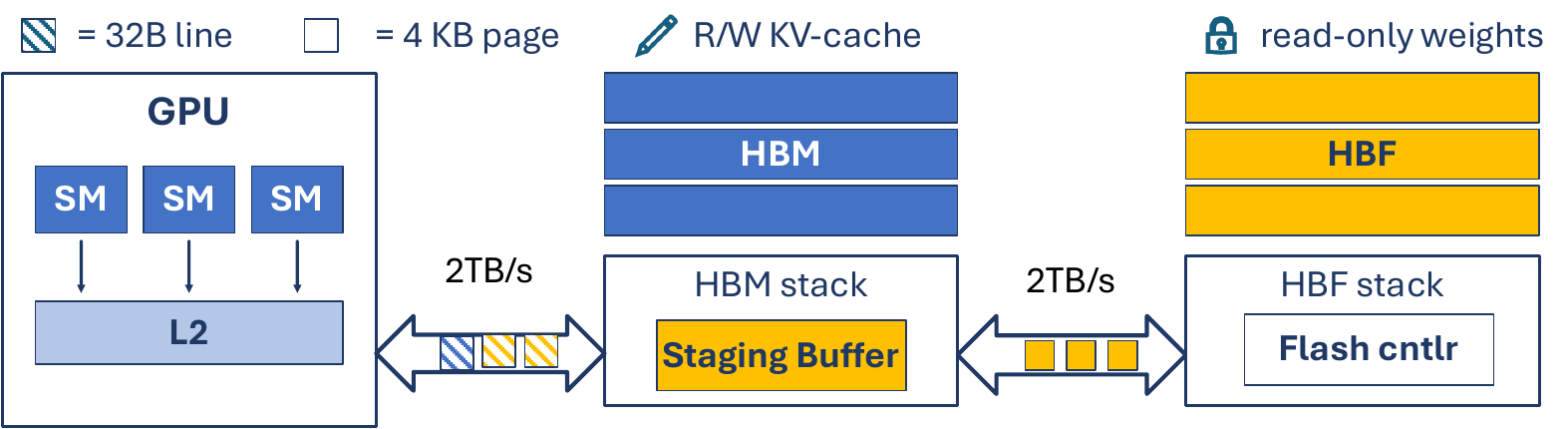}
\caption{H$^3$-like system organization~\cite{h3_llm}: HBF stores read-only
weights, HBM stores mutable data, and a staging buffer hides HBF
latency from SMs.}
  \label{fig:hbf_substrate}
\end{figure}

Figure~\ref{fig:hbf_substrate} shows our
baseline system organization, which follows $H^3$, the first
system-architecture proposal to employ HBF~\cite{h3_llm}: read-only
weights reside in HBF, mutable activations and KV state remain in HBM, and
a staging buffer helps hide the access latency.
\revhl{Each HBF stack connects to the GPU through the base die of
an HBM stack, and the staging buffer resides on that base die. Our system pairs
five HBM stacks with five HBF stacks, so each base die holds an 8\,MB slice of the 40\,MB buffer\footnote{\revhl{HBF pages
reside within a stack at 4\,KB granularity. The address translation layer
routes the prefetch and demand loads of HBF pages to the corresponding stack. Page placement is static and managed by the OS.}}.}
HBF data is not staged at HBM.

\subsection{Simulation Framework}
\label{sec:sim_method}

We use Accel-Sim~v1.3~\cite{khairy2020accel}, a cycle-level GPU simulator,
interfaced with Ramulator2~\cite{luo2023ramulator}, a memory-system simulator to serve HBM requests.
\revhl{We extend memory system of Ramulator with an HBF model} and
route read-only weight pages to HBF and all other traffic to HBM.
\revhl{Each 4\,KB page fetch incurs an unloaded request-to-return latency of
20\,$\mu$s, and additional queuing delays. HBF page fetches wait for HBF channel bandwidth and
for an install slot in the staging buffer (16-banks), and cacheline requests wait
for lookup and fill ports of the buffer.} The Accel-Sim setup matches A100
silicon with a 0.7\% MAPE on bandwidth-intensive streaming kernels. The modeled
GPU provides 80\,GB of HBM in the baseline, and 1{,}280\,GB of HBF (16$\times$)
at equal aggregate read bandwidth in the HBF system. \revhl{\Cref{sec:tact_latency_sens} further
evaluates sensitivity to flash architecture-induced latency variance~\cite{ldpc,ldpc_ssd,ldpc_retry}.}

Simulations report execution time normalized to an idealized HBF system with
the same latency as HBM. 
\revhl{Ideal-HBF runs 2.9\% faster than
a hypothetical HBM-based system iso-capacity to HBF, though ideal-HBF is not a strict upper bound.}
Real-GPU runs report runtime and resource
overheads. \Cref{tab:method_config} provides the system
configuration.

\begin{table}[!hb]
  \centering
  \caption{Configuration of the baseline HBM-HBF system.}
  \label{tab:method_config}
  \begin{footnotesize}
  \begin{tabular}{lr}
    \toprule
    \textbf{Parameter} & \textbf{Value} \\
    \midrule
    SM count & 108 \\
    GPU core / memory clock & 1.410 / 1.512\,GHz \\
    GPU, HBM Cacheline size & 128B, 32B \\
    L1 / shared memory per SM & 192\,KB \\
    L2 capacity & 40\,MB \\
    Memory channels & 40 \\
    HBM, HBF peak read bandwidth & 2TB/s, 2TB/s \\
    Aggregate HBM, HBF capacity  & 80\,GB, 1280\,GB \\
    HBF page size / request-to-return latency & 4\,KB / 20\,$\mu$s \\
    \bottomrule
  \end{tabular}
  \end{footnotesize}
\end{table}

\subsection{Simulated Workloads}

We trace seven dense FP16 models at batch size 64:
Llama-8B, Mistral-24B, Qwen2.5-32B, DeepSeek-67B, Llama-70B, Qwen2.5-72B,
and Llama-405B \revhl{(and two MoEs in \cref{sec:moe_support})}, using NVBit~1.7.6~\cite{villa2019nvbit}.  
We simulate each operator using its traced grid, CTA resource allocation, occupancy, and instructions~\cite{cublas};
only additional staging-warp resources described in
\Cref{sec:warp} are added.
\Cref{tab:method_workloads}
characterizes the weight-reading kernels (other operators like LayerNorm are fused~\cite{zhuang2024mononn, nayak2024fusemax, chimera}). 
Execution time scales with model size
but CTAs launched within each kernel remain similar. 
For example,
a Llama-405B layer reads 15$\times$ more weight pages
than Llama-8B, but the QKV projection kernel only launches 6\% more CTAs.

\subsection{Execution Model of \papername}

\papername operates with conventional GPU kernels.
Each workload CTA contains its original work warps
and one staging warp. At kernel entry, the staging warp begins
replaying that CTA's profiled page stream while the work warps
execute the operator. Because \texttt{STAGE} is nonblocking
and requests one 4\,KB page, the staging warp rapidly
establishes sufficient lookahead within the kernel. 
\revhl{Neither \papername nor H$^3$ cross kernel boundaries.}
Consequently, each weight-reading kernel incurs a startup
interval before the staging stream reaches its steady-state
lead, and all reported results include this cold-start latency.
\revhl{\Cref{sec:cross_kernel_staging} discusses how to remove this cold start.}

\section{\papername: HW/SW Co-Designed Prefetching}
\label{sec:design}

We now present \emph{\papername}\footnote{\papername takes its name
from the reduced-drag region behind a lead racer that helps a close
follower move efficiently. Similarly, \papername stages data ahead of
demand, reducing the effective latency of demand accesses.}, a
software--hardware co-designed approach to hide HBF latency.
This section describes the software component and \Cref{sec:slipstream_s} covers architectural support.

\subsection{Profiling and Kernel Modifications}
\label{sec:profile_modify}

\begin{figure}[htb]
  \centering
  \includegraphics[width=\columnwidth]{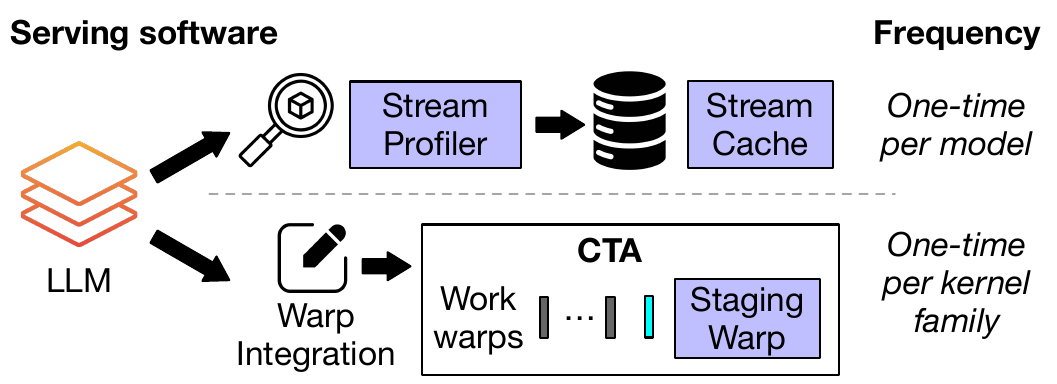}
    \caption{Overview of \papername profiler. The stream profiler collects access streams and stores them in a reusable stream cache database. We integrate a staging warp per CTA of the workload to use the access streams at runtime.}
  \label{fig:profiler_overview}
\end{figure}

We observe that LLMs repeatedly execute the same kernels with the same launch shapes across decoder layers/steps. For data-independent kernels, this repetition produces the same per-CTA page-access order. 
Consequently, when the serving system first hosts a given model and kernel-library configuration, a \textit{Stream Profiler} traces one representative execution and stores each CTA's tensor-page sequence---its \textit{access stream}---in a compact and reusable
\textit{Stream Cache}. \papername then augments each workload CTA with a \textit{staging warp} that replays the access stream at runtime, staging HBF pages into the staging buffer ahead of their demand accesses. 
\Cref{fig:profiler_overview} provides an overview of the \papername-SW framework.
\revhl{
We focus on data-independent kernels in
\cref{sec:design}--\cref{sec:evaluation} and \cref{sec:moe_support} describes the lightweight changes to provide support for data-dependent kernels (e.g., expert kernels in MoE models).
}

\subsubsection{Stream Profiler}
\label{sec:profiler}

The \textit{Stream Profiler} comprises a \textit{tracer}, which records workload CTAs' page accesses, and a \textit{validator}, which confirms that the resulting access streams remain  safe to reuse during subsequent executions at runtime.

\parhead{Tracer}
We use binary instrumentation (e.g., NVBit) to record the virtual
address (VA) of every global load, together with the kernel identity,
launch configuration, tensor sizes, and issuing CTA. Because VAs may
change across runs, the tracer maps each weight address to a
tensor-relative page index and emits a
(\emph{tensor},~\emph{page-index}) tuple. Loads outside the input
weight tensors, such as activations and KV-cache accesses, are excluded.
The tracer records each kernel invocation as a separate set of per-CTA
page sequences, with kernel-launch boundaries delimiting the profiled
streams.

\parhead{Validator} 
\papername runs the tracer across multiple inputs, and the validator admits a stream only if the per-CTA access orders match exactly, indicating data-independence. The validator also assumes consume-once layout (\Cref{sec:hbf_substrate}) and may repack weights in tile-major layout when necessary. If a tensor still fails the data-independence check, the profiler directs the serving system to retain that tensor in HBM.
This ensures \papername has accurate access streams for all validated kernels.
In practice, for dense models, only the token-embedding table---a data-dependent gather---and small normalization and bias tensors fail this check: on Qwen2.5-32B, they account for 8\% of weight bytes but 0.1\% of weight decode traffic.

\subsubsection{Stream Cache}
\label{sec:cache}
Once generated, the \textit{Stream Cache} serves as a reusable database of profiled LLM kernel families and their access streams. The cache stores one stream per kernel \emph{class} rather than one per LLM layer, which substantially reduces its footprint. We observe that demand order of LLM kernels is determined by their launch shape and tile schedule. 
We therefore key each stream by kernel name, launch parameters (e.g., grid and block dimensions), and tensor sizes (thus, no virtual addresses). Layers sharing a kernel class differ only in the weight tensors they access and the \emph{binder}  (\Cref{sec:binder}) resolves these identities at model load time for dense LLMs. For example, five stream classes totaling 1.6\,MB cover the 257 weight-operator bindings of Llama-8B for a given input shape (0.03\% of the model weights).
Because entries are keyed by launch parameters, serving the same model at multiple batch sizes adds one entry per distinct launch shape, at negligible storage cost.
The Stream Cache resides on disk and is reusable across GPU drivers and even GPU architectures, provided kernel parameters and tensor sizes are unchanged at runtime.

\subsubsection{Staging Warp}
\label{sec:warp}
The GPU assigns CTAs to SMs only at runtime. \papername therefore adds one
\textit{staging warp} to each CTA.  The staging
warp reads its assigned stream from HBM, converts each tensor-relative
stream entry into a virtual address, and issues a \texttt{STAGE}
instruction, our new ISA extension. Each \texttt{STAGE} instruction specifies a 4\,KB-aligned virtual address and
issues a non-blocking request for the corresponding 4\,KB weight page, which
HBF returns into the staging buffer ahead of demand.

We integrate the staging warp once per kernel family, independent of model or GPU: an entry macro in the serving stack's kernel-library template redirects the added warp to a generic replay routine, and each CTA's launch configuration grows by one warp. This costs $\sim$40 lines per kernel family, after which any model using that family runs on \papername unmodified. We integrated Marlin~\cite{frantar2025marlin} once with Llama-8B, and the remaining six evaluated models required no changes.

The staging warp competes with the work warps for SM resources: a warp
slot, registers, and instruction fetch and issue bandwidth. We model this
contention explicitly: in simulation, the staging warp is an additional
resident warp with its own register allocation, warp-slot occupancy, and
fetch/issue traffic; on real GPUs, we emulate it as an added warp that
issues L2 prefetches. \revhl{Staging warp execution is low-cost. It issues one \texttt{STAGE} per
4\,KB page, occupies under 0.5\% of issue slots, and runs at least 20$\mu$s ahead of work warps. The warp scheduler runs unmodified (i.e., staging warps are not given any priority).} 
Warp-specialized kernels with full register occupancy (e.g., CUTLASS~3.x)
still accommodate the replay routine: we fold it into an idle warp of an
existing warpgroup (e.g., a spare TMA producer warp), adding no SM
resources. 

\subsection{Bootstrapping Access Streams to Warps}
\label{sec:binder}

\begin{figure}[htb]
  \centering
  \includegraphics[width=\columnwidth]{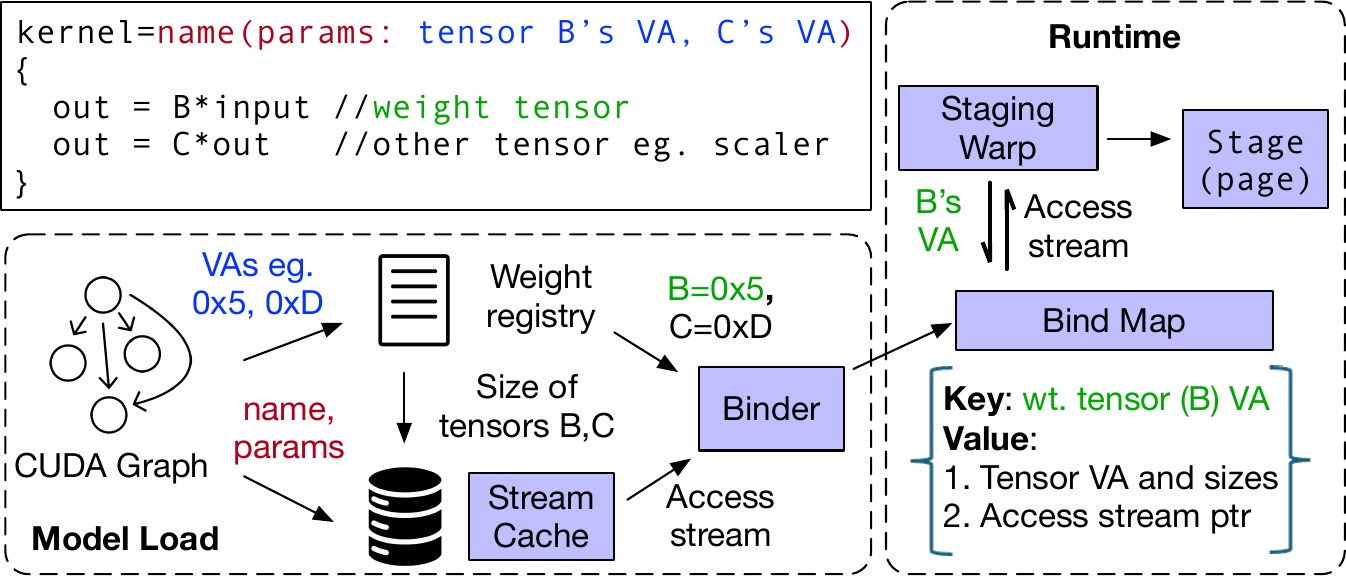}
    \caption{\papername bootstrapping. The binder gets kernel parameters from CUDA graph capture. It then uses the Stream Cache and serving system's weight registry to build the bind map which staging warps use at runtime.}
  \label{fig:binder_overview}
\end{figure}

The Stream Cache contains tensor-relative page indices, whereas a
\texttt{STAGE} instruction requires a virtual address that becomes known
only after the serving system loads the model. Furthermore, because the
workload kernel has already been compiled and captured in the decode
CUDA graph, the staging warp cannot receive a profile-specific stream
pointer as an additional launch argument. \papername bridges this gap using a
\textit{binder}. The binder exploits the weight tensor's base address,
which is available both in the captured graph node and as a runtime
kernel argument, as a common key between the serving software and the
staging warp. \Cref{fig:binder_overview} shows the overview of \papername's bootstrapping.

At model load, the binder 
walks the weight-kernel nodes in the captured CUDA graphs and matches
each node to an entry in the \textit{Stream Cache} using its kernel
parameters. It then creates a \textit{Bind Map} entry keyed by the
kernel's weight-tensor base address. Each entry stores the base addresses
of all tensors accessed by the operator, such as weights and scales,
along with a pointer to the corresponding access stream. At runtime, the
staging warp uses the weight pointer already passed to the kernel to
locate this entry, reconstruct the virtual addresses from the
tensor-relative page indices, and select the portion of the stream
associated with its CTA. For Llama-8B, the binder reads
1.6\,MB of access streams and creates 14\,KB of \textit{Bind Map}
entries.
Because addresses are rebound at each model load, the same stream survives
 process restarts.

\subsection{\papername Software's Evaluation Methodology}
\label{sec:method_software}

We evaluate \papername's software stack and staging-warp overhead on
A100 (108 SMs), H100 (132 SMs), and Blackwell RTX~PRO~6000
(188 SMs) GPUs. All seven models of \Cref{tab:method_workloads} run  GPTQ-INT4 with Marlin kernels~\cite{frantar2025marlin}.
\revhl{We use GPTQ-INT4 because the larger FP16 models (up to 800\,GB)
exceed the memory of our GPUs even when profiling few layers.}
We also evaluate W8A8 Llama-8B and the FP16 LM-head (both CUTLASS~\cite{tiling_cutlass}). 
\revhl{Overall, we profile 84 kernel classes---74 Marlin, 8 W8A8, and 2 FP16---through vLLM~\cite{vllm}
with CUDA-graph decoding at batch size~8 (for low KV pressure)}. Because \texttt{STAGE} is
unavailable on current GPUs, we measure replay overhead by adding a
resident staging warp that issues L2 prefetches in the recorded page
order.
For the largest models we profile a subset
of layers: 20/80 for Llama-70B and Qwen2.5-72B, 24/95 for DeepSeek-67B, and
8/126 for Llama-405B. We report full-model equivalents (e.g., Llama-70B's
embedding-table overhead shrinks 4$\times$ at 80 layers).

\subsection{Results: Overhead of Staging Warps}
\label{sec:tact_touch_results}

\begin{figure}[htb]
  \centering
  \includegraphics[width=\columnwidth]{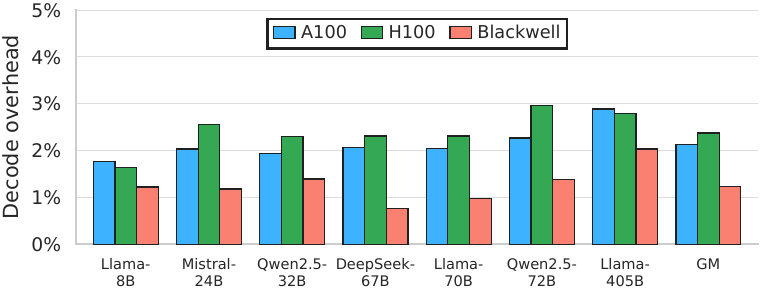}
  \caption{Real-GPU decode overhead of replaying \papername's access
  streams.}
  \label{fig:tact_replay_overhead}
\end{figure}

The staging warp incurs low overhead on GPUs.
\Cref{fig:tact_replay_overhead} reports 0.76--2.96\% end-to-end decode
overhead across the seven models (geometric means 2.1\%, 2.4\%, and
1.2\% on A100, H100, and Blackwell). 
\revhl{At kernel launch on A100, we measure that the Bind Map lookup adds a nominal 0.15\,$\mu$s latency.
Moreover, 128B cachelines of the access stream allows staging 128KB of data for 0.1\% extra traffic.
Our real-silicon and simulated results account for these costs (\cref{sec:tact_eval}).} 
The warp uses no dynamic shared memory and adds 2--4 registers per
thread, but the added warp grows a CTA from 256 to 288 threads and
3.6\,K registers, 5.5\% of the 64\,K register file, while occupancy remains unchanged (\Cref{sec:warp}).

\subsection{Results: Stream Cache Coverage Analysis}
\label{sec:tact_coverage}

\subsubsection{Kernel class coverage}
The offline
validation reproduces the exact per-CTA page order for \textit{all}
evaluated kernel classes, with generated tokens remaining bit-identical
on all three architectures. 
Four stock W8A8 layout
revisits each page 28.6--31$\times$. After \papername validator's repacking, all four evaluated
W8A8 classes satisfy the consume-once requirement. 

\subsubsection{Weight coverage}
Validated streams cover nearly all measured weight traffic, with 99.1--99.95\% traffic coverage
(geometric mean 99.6\%); the remaining traffic comprises normalization
parameters, biases, and group indices, which account for at most 0.9\%
of weight traffic and \papername retains in HBM. Weight-storage
coverage is 81.0--97.2\% (geometric mean 91.9\%). The embedding
table---a data-dependent gather---is the
only HBM-resident weight, and it scales with vocabulary rather than
model depth: the highest loss of coverage is for Llama-8B at 19\% of model weights, while Llama-405B loses just 2.8\% coverage.

\subsubsection{Profiling overheads}
Offline profiling
takes 40 minutes to 2.6 hours per model, while model-load binding takes
approximately one minute. Moreover, a stream acquired on one GPU
generation can be reused across all three evaluated generations, so
\papername requires neither manual kernel analysis nor repeated
deployment-time profiling.

\section{Architectural Support for \papername}
\label{sec:slipstream_s}

\subsection{\papername Architecture Overview}%

\begin{figure}[htb]
  \centering
  \includegraphics[width=\columnwidth]{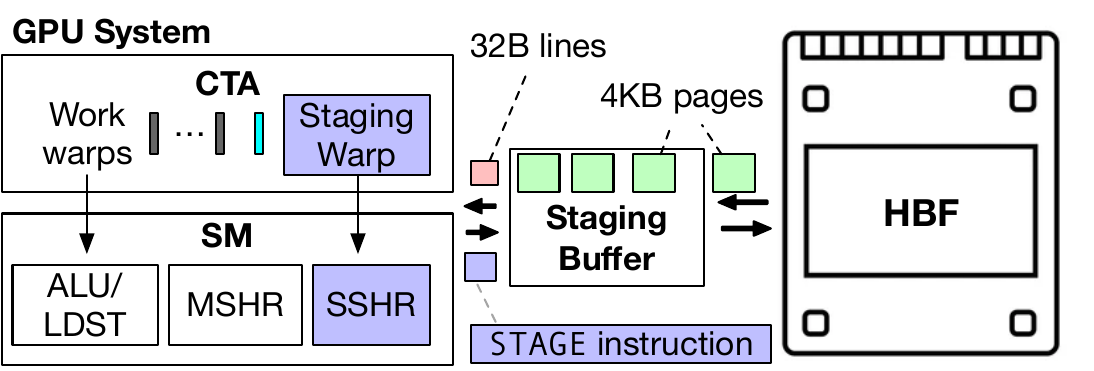}
    \caption{Overview of \papername's architecture. Staging warps issue
    \texttt{STAGE} instructions, the per-SM SSHR tracks each staged page
    until its first demand access, staging buffer serves demands at near-HBM bandwidth/latency.
    }
  \label{fig:arch_overview}
\end{figure}

\papername's software pipeline inserts \texttt{STAGE} instructions that move pages from HBF ahead of demand; we now describe the architectural
support that executes them. Each \texttt{STAGE} prefetches a 4\,KB page
into the \textit{staging buffer}, from which subsequent demand loads are served at
near-HBM latency. 
While we employ a dedicated buffer, in principle on-chip GPU caches can also be repurposed as the staging buffer (especially for \papername 16MB owing to its small buffer)~\cite{selfa2017application}.
Because \texttt{STAGE} instructions generate long-latency memory
requests, \papername adds a per-SM \textit{Stage Status Holding Register} (SSHR) to
track their progress. 
\Cref{fig:arch_overview} presents these architectural components, which
we describe next.

\subsection{Stage Status Holding Register (SSHR)}
\label{sec:opt}

The per-SM SSHR tracks \texttt{STAGE} requests and bounds how far the
staging stream can run ahead of demand. Each \texttt{STAGE} instruction
allocates one SSHR entry for a 4\,KB page. Demand requests consult the
SSHR before memory system access. If a demand matches an in-flight page entry, the requesting warp waits on that entry,
analogous to a miss waiting on an MSHR.

\papername employs a \emph{cooperative page-access protocol}: once the
page is installed in the staging buffer, the buffer returns an
\texttt{ACK} to requesting SM, and SSHR marks the corresponding
entry as ready. Any warps waiting on the corresponding entry issue their
cache-line demand requests to the memory system. A demand load that finds
no SSHR entry issues to the memory system unchanged: staging is a
performance hint, not a correctness requirement. Like ordinary loads,
\texttt{STAGE} requests carry virtual addresses and translate through the GPU TLB hierarchy.
\revhl{\texttt{STAGE}s and \texttt{ACK}s (8B flits) use the existing on-chip
network (0.8\% extra HBF traffic). The network routes each \texttt{STAGE} by page
address to the memory partition that owns it. Demand loads proceed unmodified.}

For timely staging of data, we want knowledge of demand's consumption rate (\Cref{sec:h3_limitations}).
A conventional MSHR releases its entry when the requested data
returns, and so MSHRs carry no information about consumption. \textbf{Our key
insight is to retain each SSHR entry until the staged page receives its
first demand access.} When all SSHR
entries are occupied, the staging warp is backpressured; each first
demand response releases one entry and admits the next \texttt{STAGE}
instruction. SSHR occupancy therefore counts the pages staged but not
yet demanded, 
so demand itself paces the staging stream and a slow
consumer cannot overrun the staging buffer. As each
entry tracks 4\,KB page rather than 128-byte cache line, \papername
reduces tracking state by $32\times$ compared to MSHRs.

We size the SSHR to provide the bandwidth--latency product of staging
requests (as required by Little's Law). In our system, $2\,\text{TB/s} \times 20\,\mu\text{s}$
corresponds to approximately 40\,MB, or 9{,}800 4\,KB pages. Distributing
these pages across all 108 SMs requires 90 entries per SM.
Some LLM operators run on fewer SMs---the sparsest measured operators
launch just 56--64 CTAs on the 108-SM A100
(\Cref{tab:method_workloads})\revhl{. We provision 256 SSHR entries
per SM, so that about one third of the SMs can saturate HBF
bandwidth when necessary. Each entry stores a 37-bit page tag, a valid bit, and
a ready bit, padded to 5\,B, so the SSHR of each SM occupies only 1.25\,KB
(\Cref{tab:tact_physical_cost}).} A per-SM \texttt{SSHRMax} register exposes only the appropriate number of entries
during each kernel based on the active-SM count (the value is stored at the prologue of access stream).
\revhl{Moreover, grid quantization gives some SMs more CTAs than others (\Cref{sec:h3_limitations}). SSHRMax is
therefore weighted by the page count of each CTA (computed by the stream
profiler) and the CTA count per SM (computed by the hardware).
Thus, SMs with more CTAs/ page-accesses receive proportionally more staging budget
than other SMs.}
This keeps the in-flight staging window saturated while staging remains demand-proportional.

\begin{figure*}[htb]
  \centering
  \includegraphics[width=\textwidth]{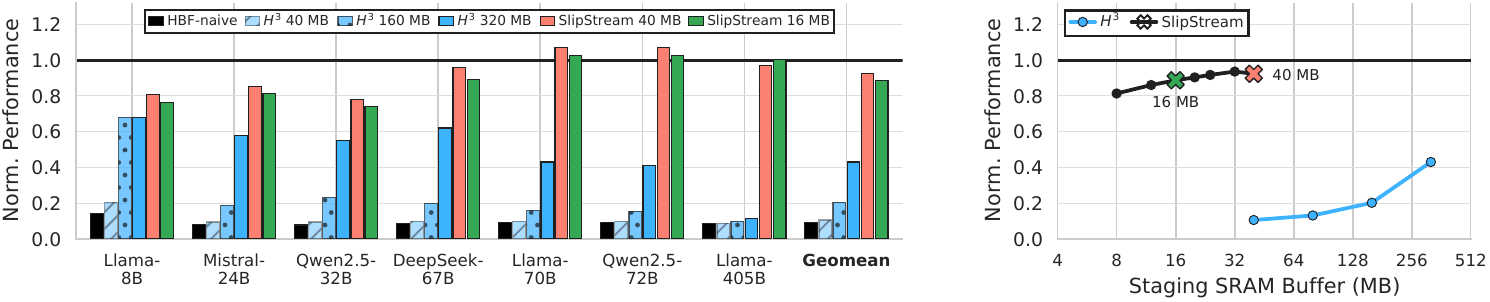}
  \caption{Performance of on-demand fetch, $H^3$, and
  \papername (40\,MB and 16\,MB buffers), normalized to an ideal-HBF
  system with HBM-like latency (left), and
  geometric-mean performance against buffer SRAM (right). \papername performs within 8\% (12\%) of ideal-HBF at just 40MB (16MB) of buffer SRAM.}
  \label{fig:tact_baseline}
\end{figure*}

\subsection{Staging Buffer}
\label{sec:staging_buffer}

The \textit{staging buffer} serves both \texttt{STAGE} and demand
requests (see~\Cref{fig:hbf_substrate}). A \texttt{STAGE} instruction
initiates 4\,KB transfer from HBF, whereas demand loads are line-granular. Demands normally arrive after the page is
installed. If a demand arrives while the page is in flight, it waits on
the SM's SSHR entry; a request from another SM to the same page
coalesces with the shared outstanding request at the buffer.
Upon completion, the buffer installs the 4\,KB page and returns an
\texttt{ACK} to the requesting SM, allowing waiting demands to proceed.

Little's Law requires
40MB of requests in
flight to sustain peak HBF bandwidth. This requirement, however, does not imply that the staging buffer
must reserve a data slot for every in-flight page. \papername therefore
tracks pages while they are in flight but allocates data SRAM only when
they return. Provisioning the staging buffer below 40\,MB
introduces two challenges: bounding how many pages return before demand
reaches them and retaining resident pages until their final use.

\parhead{Demand-paced admission.}
Demand consumption rate is slow at kernel launch (e.g., due to prologue operations).
As such, staging warps could populate the entire
SSHR window before the new kernel issues any
demand requests, causing more pages to return than a smaller-than-40MB buffer can hold.
\papername therefore begins each kernel with an initial staging
allowance, $S_0$, whose aggregate value does not exceed the staging buffer
capacity. Each SM tracks $U$, the number of staged pages that have not
yet received their first demand, including both in-flight and resident
pages, and $F$, the cumulative number of first demands in the current
kernel. The SSHR admits a \texttt{STAGE} instruction if $U < \min(\texttt{SSHRMax},\,S_0 + F).$

Each first demand decreases $U$ and increments $F$, allowing the staging
window to grow toward \texttt{SSHRMax} as demand advances. The counters
are reset at each kernel launch, and $S_0$ is recomputed from the buffer capacity and active-SM count.

\parhead{Consumption-aware eviction.}
A smaller buffer may evict
a page before demand loads have consumed all of its cache lines. 

\noindent The
consume-once layout guarantees that every staged page is read in full,
allowing a per-page completion counter to identify its final demand.
\revhl{
The counter counts the consume-once read; later consumers either hit in L2 or re-STAGE the page, which coalesces if the page is in flight or hits if resident.}
\papername marks a page evictable only after this final access and
appends it to a consumed-page queue. A prefetched page installs into a
free slot or replaces the queue head, ensuring residence of partially consumed pages.
\revhl{Moreover, to maximize shared pages' residence, the buffer prioritizes unshared pages for eviction. Pages with very high reuse distance can still be refetched via subsequent \texttt{STAGE} instructions.}

\revhl{If a page returns when no buffer slot is available, the buffer
discards the payload, marks the request retry-pending using a single bit in its tag (a discarded page retains its tag entry), and reissues the
HBF access}. Newly reclaimed slots are reserved for retries before
younger arrivals. If a retry still cannot obtain a slot, it preempts the
youngest resident page that has not yet received demand, returning that
page to the staging pipeline. Together, they guarantee progress without
evicting partially consumed data. Thus, first demand regulates kernel-launch staging, while final demand regulates buffer reclamation.

We evaluate \papername variants with 40MB and 16MB buffer sizes.
For the 40MB and 16MB variants, $S_0$ value is $9800/108=90$ and $4000/108=37$, respectively.

\section{Evaluation Results}
\label{sec:evaluation}

\subsection{\papername Performance}
\label{sec:tact_eval}

\papername performs close to ideal-HBF for all LLMs. 
\Cref{fig:tact_baseline} shows the
performance of HBF-naive, $H^3$, and \papername relative to an ideal-HBF
system. 
HBF-naive reaches only 9.2\% of ideal-HBF. With a small,
model-independent 40\,MB buffer, \papername reaches 92.4\% overall and
78.0--107.2\% per model.

Llama-70B and Qwen2.5-72B exceed ideal-HBF. Both systems serve every demand
load from buffer SRAM, so they differ only in when weight traffic enters the
memory hierarchy. Ideal-HBF releases each sub-kernel's loads as an unmetered
burst at the barrier, saturating the L2 MSHR pool and stalling later requests
at the L2. \papername's pacing avoids this, reducing weight AMAT by 28\% and
L2 MSHR reservation retries by 2.45$\times$, for up to 7.2\% speedup.

Small models (8B, 24B, 32B) lose performance because the cold-start HBF
latency at each weight-kernel boundary is model-independent: it consumes 18-22\%
of $353$--$773\,\mu$s layer, but only 4\% of Llama-70B's
$1391\,\mu$s layer. \revhl{Mistral-24B has load imbalance in some kernels
(1.27$\times$ of mean pages for some SMs), which SSHRMax
mitigates.}
\revhl{Optimizing access stream reads can also reduce costs: Incurring no stream read cost improves \papername by 3.2\% (Llama-8B).}

At 16\,MB, \papername reaches 88.8\% overall and 74.1--102.7\% per model, a
4.0\% loss relative to 40\,MB. The gap comes from more conservative staging at
kernel launch: the 16\,MB variant issues just 37 pages per SM, 40\% of the
bandwidth--delay product, so kernels start more slowly. 

$H^3$'s performance is much worse than \papername. At 40\,MB it reaches only 10.6\% of ideal-HBF
performance---$4.8$--$11.5\times$ slowdowns, barely better than on-demand
fetching. Even with an impractical 320\,MB buffer, $H^3$
reaches 43.1\%. \Cref{fig:tact_baseline} (right) sweeps \papername from 8\,MB to 40\,MB:
performance degrades smoothly from 92.4\% at 40MB to 81.4\% at 8\,MB,
making \papername a scalable and low-cost solution.
\revhl{Even at 8MB buffer, only Qwen2.5-32B discards pages (\checkvalue{0.3\%} extra HBF traffic). No other model/ buffer-size incur discards.
}

\subsection{Analysis of HBF-Induced Stalls}
\label{sec:tact_stalls}

\begin{figure}[htb]
  \centering
  \includegraphics[width=\columnwidth]{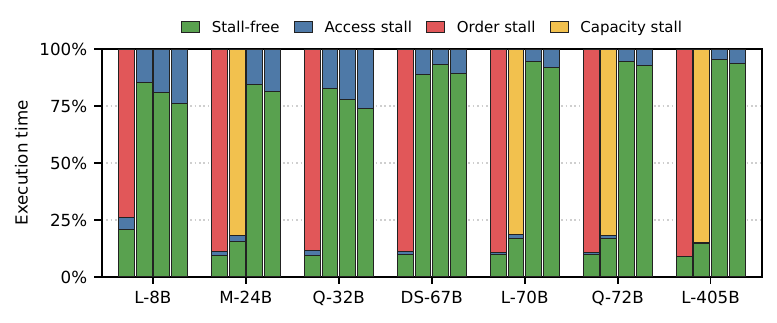}
  \caption{\revhl{Stall attribution at runtime. Bars per model, left to right: $H^3$-40MB (VA order), $H^3$-40MB
  with the oracle demand order, \papername 40\,MB, and \papername 16\,MB.
  \papername eliminates order and capacity stalls.}}
  \label{fig:tact_demand_outcomes}
\end{figure}

\revhl{\Cref{fig:tact_demand_outcomes} attributes runtime to wait-free demand loads and stall types (\Cref{sec:h3_limitations}). In \papername, loads have 100\% buffer hits (median AMAT of 0.5--1.4\,$\mu$s;
\Cref{sec:tacts_residency}), so its only stalls are access stalls, primarily
 cold starts (15--22\% of time on the small models at 40\,MB).
At 40\,MB, order stalls take 74--91\% of $H^3$'s time.
Oracle demand order (second bar) removes order stalls but exposes capacity
stalls: three models recover to 1.13--1.21$\times$ yet still incur the cold start
(11--17\% of time), while the large models (and 24B, owing to SM load
imbalance) lose 82--85\% of their time to capacity stalls and run at
5.9--6.8$\times$.
Demand-order $H^3$ runs small models slightly faster than \papername because our $H^3$ model is optimistic. Its demand loads read pages directly from fill queues of the buffer, and it provisions 2.8$\times$ more MSHRs. Under identical hardware, the two designs match for Llama-8B (within 1\%); \papername's flow control is essential at larger models/ smaller buffers.
}

\subsection{Analysis of HBF Bandwidth Utilization}
\label{sec:tact_bw}

\begin{figure}[htb]
  \centering
  \includegraphics[width=\columnwidth]{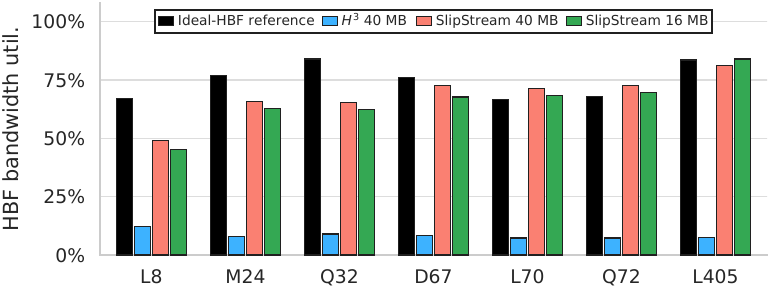}
  \caption{Sustained HBF read bandwidth (norm. to 2TB/s) for different designs. \papername retains most  bandwidth utilization in all evaluated models.}
  \label{fig:tact_bw_util}
\end{figure}

We compare each design against ideal-HBF system
 and report HBF's read bandwidth utilization in \cref{fig:tact_bw_util}.
\papername-40MB sustains 49.2–81.2\% of peak: within 2.4--3.3\% of
peak of the reference on DeepSeek-67B and Llama-405B, \emph{above} it
on Llama-70B and Qwen2.5-72B ($\approx$107\%; see \Cref{sec:tact_eval}), 
and 11.3--18.5\% below it on the
three smaller models, consistent with their performance
losses. The 16\,MB configuration stays within 5.0\% of peak of 40\,MB
on every model and is 2.8\% \emph{higher} on Llama-405B. 
In contrast, HBF-naive sustains only 6.0--7.1\%, and
$H^3$ 7.3--11.4\% at the same 40\,MB, because insufficient buffer capacity
limits in-flight traffic and thereby HBF bandwidth.

\subsection{Analysis of Buffer Residency and Load-to-use Latency}
\label{sec:tacts_residency}

\begin{figure}[htb]
  \centering
  \includegraphics[width=\columnwidth]{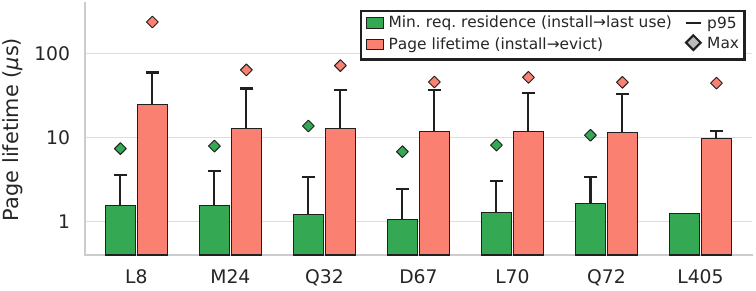}
  \caption{Per-page lifetime in the 16\,MB buffer (log scale): the
  time \papername needs each page versus the time it holds it.  The
  gap is post-consumption retention; measured to be 8--23$\mu$s (so sufficient margin against jitters).}
  \label{fig:tact_residence}
\end{figure}

\papername's buffer spends most of its capacity on retention of in-use data.
\Cref{fig:tact_residence} contrasts useful lifetime (installation to
final demand) with residence (installation to eviction) at 16\,MB.
Demand consumes a page in 1.1--1.7\,$\mu$s on average (95th percentile
below 4\,$\mu$s), but the buffer retains it for
9.8--24.7\,$\mu$s, leaving 8.5--23\,$\mu$s of margin which aids in
maintaining performance even in presence of SM load imbalance
 that can occur due to runtime congestion.
\papername does not evict pages before final use and we also
measure zero discards of pages returned by HBF.
The 16\,MB point is tightly provisioned: its maximum
install-to-first-demand lead (6.5--13.4\,$\mu$s) approaches the
$\approx$12\,$\mu$s measured average buffer residence. 
On average, though, only 8.9\% of buffer's page are staged but unconsumed,
so demand tightly follows staging.
Yet load-to-use demand latency does not increase:
median weight-load latency is 0.5--1.4\,$\mu$s and the 99th percentile
is 1.8--2.7\,$\mu$s, comparable to HBM's
0.9--1.8\,$\mu$s AMAT.

\subsection{Performance Contribution Breakdown}
\label{sec:tacts_results}
\label{sec:contribution_breakdown}

\begin{figure}[htb]
  \centering
  \includegraphics[width=\columnwidth]{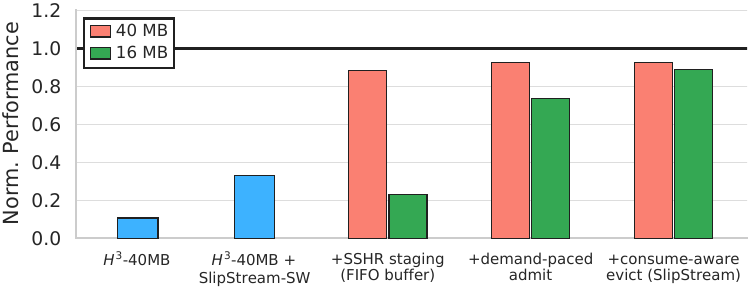}
  \caption{Contribution breakdown normalized to ideal-HBF's performance. Each optimization is necessary (especially for 16MB buffer) and additive.}
  \label{fig:tact_breakdown}
\end{figure}

\Cref{fig:tact_breakdown} shows \papername's ablated mechanisms at both buffer capacities. 
At 40\,MB, \emph{SlipStream-SW} supplies the
profiled demand order to $H^3$ but reaches only 32.8\% of ideal-HBF
performance as demand can outrun prefetch, leaving large models capacity-stalled
(\Cref{sec:tact_stalls}). SM-side SSHR staging provides the largest gain,
reaching 88.1\% with 40MB buffer. The demand-paced admission at kernel launch raises performance to 92.4\%
and makes 16MB buffer viable at 73.7\% performance of ideal-HBF.
While consumption-aware eviction has little effect for 40MB buffer owing to its size,
it is critical in improving performance of 16MB buffer, taking it to 88.8\%.
Thus, at 16\,MB, all three mechanisms become necessary. 

\subsection{Sensitivity to HBF Read Latency}
\label{sec:tact_latency_sens}

\begin{figure}[htb]
  \centering
  \includegraphics[width=\columnwidth]{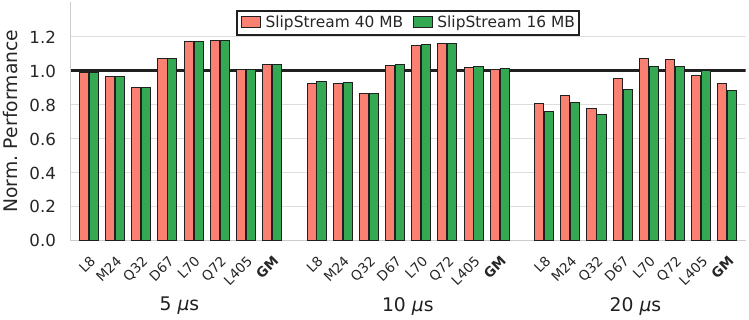}
  \caption{Sensitivity to HBF request-to-return latency, normalized to
  ideal-HBF.  \papername retains low  average overheads at various HBF latencies.}
  \label{fig:tact_latency_sens}
\end{figure}

\parhead{Unloaded Latency.}
We evaluate HBF request-to-return latencies of 5, 10, and
20\,$\mu$s (\Cref{fig:tact_latency_sens}). Both buffer
configurations stay at their design capacity, while the SSHR
window scales with the bandwidth--delay product. 
\papername
improves as latency decreases, retaining 92.4\%, 100.7\%,
and 103.7\% at 40\,MB and 88.8\%, 101.2\%, and 103.7\% at
16\,MB for 20, 10, and 5\,$\mu$s. Below 20\,$\mu$s it exceeds
ideal-HBF on average, with four of seven models individually
above it, due to the pacing reason of \Cref{sec:tact_eval}. 
The cost of a smaller 16MB buffer also reduces: the two capacities differ by
4.0\% performance at $20\mu$s, by only 0.5\% at 10\,$\mu$s, and produce
identical cycle counts on all seven models at 5\,$\mu$s, since
a shorter latency leaves fewer pages in flight and 16\,MB
already covers that window. The SSHR therefore
serves the entire 5--20\,$\mu$s range.

\parhead{\revhl{Latency Variance.}}
\revhl{
Flash adds two sources of latency variance. First, flash organizes each stack into channels and subarrays, and each read occupies a subarray for its sensing time ($t_R$), leading to subarray contention. Second, when a read has bit errors (at probability $p_e$), the controller rereads the page at other reference voltages and decodes it again, repeating across decoding levels as needed\footnote{\revhl{Write interference, a third common source, is virtually absent because weights are read-only; only periodic read-disturb refresh rewrites them.}}. Vendors have not disclosed these parameters for HBF, and recent HBF studies assume a fixed read latency~\cite{h3_llm, hbf_spec, hbf_explore, hbf_flashaccel}. We add both effects to the 20\,$\mu$s unloaded latency. We model the subarray organization of~\cite{hsu2026haven} with $t_R$ of 3 and 5\,$\mu$s~\cite{cheong2018flash, xlflash}, and set $p_e$ to $10^{-4}$~\cite{ldpc}; each further decoding level is 100$\times$ less likely and adds $1.75\,t_R$. Subarray contention accounts for most of the loss, mainly at kernel launch, when the initial burst of reads contends on subarrays. Read-retry costs at most one point, because retries are rare. Together, the two effects lower \papername from 92.4\% to 89.2\% and 86.9\% of ideal-HBF at 40\,MB for $t_R$ of 3 and 5\,$\mu$s, and from 88.8\% to 87.9\% and 84.8\% at 16\,MB. Neither effect causes page discards or buffer misses.
}

  \subsection{Storage, Area, and Power Overheads}
  \label{sec:tact_hardware_cost}

\Cref{tab:tact_physical_cost} reports CACTI~\cite{thoziyoor2008cacti} estimates at 22\,nm for the structures that \papername adds.
\revhl{On the GPU die, \papername adds only the SSHRs and the SSHRMax registers. Each SM searches its 1.25\,KB SSHR in one cycle, and all 108 SSHRs occupy 0.39\,mm$^2$, or 0.02\% of the 826\,mm$^2$ GA100 die when scaled to 7\,nm.}
\revhl{The staging buffer and its tags reside on the five HBM base dies (\Cref{sec:sys_org}). Each 8\,MB slice uses 16 banks to sustain 0.4\,TB/s of installs from flash alongside 0.4\,TB/s of demand reads. The slice occupies 6.3\% of a 121\,mm$^2$ base die at 22\,nm, or 1.9\% at the 3\,nm bitcell that $H^3$ assumes~\cite{h3_llm}. $H^3$-320MB occupies 42.7\% and 12.7\%, respectively (10.7\% by $H^3$'s bitcell-only formula). The tag array scales with bandwidth--delay product rather than with buffer capacity. The 2.5\,ns buffer access is negligible against HBF's 20\,$\mu$s latency.}

\revhl{The five base dies leak 12.9\,W and draw at most 23.3\,W of dynamic power. At the measured 49.2--81.2\% HBF utilization, \papername therefore adds 24.5--32.0\,W. The HBF stacks draw 4$\times$ the energy per bit of HBM, which raises the A100 envelope from 400\,W to 654\,W. \papername adds 3.7--4.9\% to this envelope at 40\,MB and 1.9--2.8\% at 16\,MB, while $H^3$ (320MB) adds a significant 18--22\% to system power.}

\begin{table}[htb]
\centering
\caption{Physical cost of \papername's added structures.
Top-cell value is for \papername-40MB, bottom is 16MB variant.}
\label{tab:tact_physical_cost}
\begin{scriptsize}
\begin{tabular}{@{}lrrrrr@{}}
  \toprule
  \textbf{Structure} & \textbf{Storage} & \textbf{Area} &
  \textbf{Access} & \textbf{Leak.} & \textbf{Dyn.} \\
   & & (mm$^2$) & (ns) & (W) & (W) \\
  \midrule
  \multicolumn{6}{@{}l}{\textit{\revhl{GPU die (per SM)}}} \\
  SSHR CAM & 1.25\,KB & 0.0036 & 0.10 & 0.001 & 0.001 \\
  SSHRMax register & 4\,B & --- & --- & --- & --- \\
  \addlinespace
  \textbf{\revhl{Total (108 SMs)}} & \textbf{135\,KB} & \textbf{0.39} & --- &
    \textbf{0.153} & \textbf{0.09} \\
  \midrule
  \multicolumn{6}{@{}l}{\textit{\revhl{HBM base die (per stack)}}} \\
  Staging buffer   & 8\,MB   & 7.62 & 2.5 & 2.56 & 4.65 \\
                   & 3.2\,MB & 3.39 & 1.6 & 1.13 & 2.95 \\
  \addlinespace
  Tag + metadata   & 32\,KB   & 0.026 & $<$0.2 & 0.010 & 0.002 \\
                   & 25.6\,KB & 0.022 & $<$0.2 & 0.008 & $\sim$0 \\
  \addlinespace
  Control + queues & 1.7\,KB & 0.028 & $<$0.2 & 0.001 & $\sim$0 \\
  \addlinespace
  \textbf{\revhl{Total (5 stacks)}} & \textbf{40.2\,MB} & \textbf{38.35} & --- &
    \textbf{12.86} & \textbf{23.26} \\
                   & \textbf{16.1\,MB} & \textbf{17.21} & --- &
    \textbf{5.69} & \textbf{14.77} \\
  \bottomrule
\end{tabular}
\end{scriptsize}
\end{table}

\section{Discussion}
\label{sec:discussion}

\subsection{\revhl{Cross-kernel Staging}}
\label{sec:cross_kernel_staging}

We can avoid the 20$\mu$s cold-start at kernel launch by letting the 
staging warp cross kernel boundaries. 
\revhl{Weights do not depend on the output of the previous kernel, so staging them
before launch is safe. The obstacle is attribution, because the staging warp
must know which SM consumes each page. Persistent
megakernels~\cite{mirage} remove this obstacle by
fixing the operation sequence of each CTA at compile time. One staging warp per
SM then persists across all operations. It begins the next operation once it has
issued its current stream, and each SM limits its next-operation pages at its
share of the buffer (no new barrier is needed). 
In our simulations of such a megakernel, cross-operation staging raises
SlipStream from 92.4\% to 96.1\% of ideal-HBF at 40\,MB and from 88.8\% to
99.5\% at 16\,MB, owing to even better pacing and less L2 congestion at kernel starts with 16MB buffer.
The worst-case model's slowdown falls from 28--35\% to just 13--16\%.}

\subsection{\revhl{Support for Other Inference Deployment Targets}}
\label{sec:moe_support}

\parhead{\revhl{Prefill.}}
\revhl{Prefill kernels are data-independent, so \papername profiles and replays them similar to decode kernels. Serving engines bucket kernel shapes by prompt tokens, and the Stream Cache holds one stream per launch shape. Compute-bound prefill also shares each weight page across multiple CTA tiles, which \papername handles natively (\cref{sec:staging_buffer}).}

\begin{figure}[htb]
  \centering
  \includegraphics[width=\columnwidth]{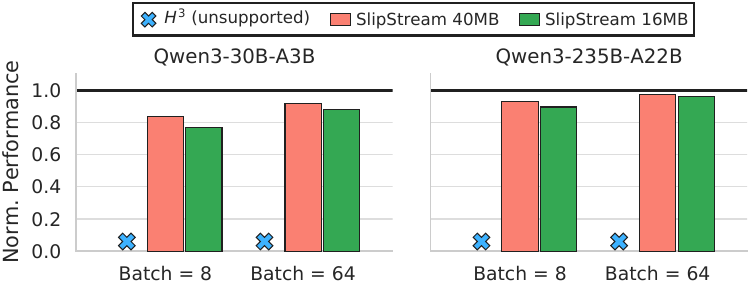}
  \caption{\revhl{Performance of MoE kernels~\cite{trtllm}, normalized to ideal-HBF. \papername-40MB reaches 91.2\% of ideal-HBF (87.3\% at 16MB).}}
  \label{fig:tact_moe}
\end{figure}

\revhl{
\parhead{Mixture-of-Experts models.} In an MoE layer~\cite{shazeer2017moe, switchtransformer, jiang2024mixtral, deepseekv3}, a \textit{router} assigns each token to a subset of the \textit{experts}, so expert weights that a kernel reads are data-dependent. We examine popular MoE kernels~\cite{vllm, trtllm} and make two observations. First, routing completes before the MoE kernel launches, and the kernel receives the routing result as an input, so experts and tiles of each CTA are known at launch. Second, all experts share one shape and tiling, so the page order within a tile is identical across experts. The access stream of each CTA is therefore known at kernel entry, without prediction.

We extend \papername-SW to support MoEs. First, the Stream Profiler records one expert-relative stream per tile, along with the tile scheduler of the kernel. The validator admits the kernel only if, across routings, every tile follows its stream and every CTA follows the scheduler\footnote{\revhl{Kernels that assign tiles dynamically (e.g., work stealing) require each CTA to claim its tiles at least one HBF latency (20\,$\mu$s) before computing them, so that the staging warp stages them in time.}}. Second, the Bind Map entry of each MoE layer adds the expert stride, from which the staging warp computes the base address of each expert. Third, at kernel entry, the staging warp runs the tile scheduler on the routing input, as the work warps do, to find the tiles of its CTA. It sets \texttt{SSHRMax} from relative load of its CTA and replays stream of each tile from base address of its expert.

We trace and simulate persistent MoE kernels of TensorRT-LLM~\cite{trtllm} for Qwen3-30B and Qwen3-235B~\cite{yang2025qwen3} (the w13 and w2 GEMMs are fused). At batch sizes of 8 and 64, 31 and 87 of 128 experts activate on average. \Cref{fig:tact_moe} shows that \papername reaches 91.2\% of ideal-HBF at 40\,MB and 87.3\% at 16\,MB (geometric mean), with no discarded pages or buffer misses. 
The bandwidth utilization with ideal-HBF is very high at 96--98\%, so \papername's pacing does not provide additional improvement. At batch 64, 6.7\% of tiles share weight pages with another tile; \papername-HW coalesces their \texttt{STAGE}s and serves the shared pages without refetches. Performance improves with batch/ model size (HBF's target regime), from 83.7\% on Qwen3-30B at batch 8 to 97.1\% on Qwen3-235B at batch 64. The fixed 20\,$\mu$s cold start per layer amortizes over kernels that grow from 173\,$\mu$s to 1.82\,ms. Even at batch=1, \papername retains 82.2\% of ideal-HBF on Qwen3-235B.
}

\parhead{\revhl{Shared-prefix KV cache.}}
\revhl{Supporting this follows the same approach as MoEs. Long-lived prefix KV is written once and read many times, which suits HBF. Paged-attention kernels locate KV pages through a block table that the serving engine writes before launch~\cite{vllm}. The staging warp therefore reads the table, as the work warps do, and stages the KV pages in walk order. Requests that share a prefix read the same pages, which \papername handles natively as shared pages (\cref{sec:staging_buffer}).}

\parhead{\revhl{Limitations.}}
\revhl{\papername does not directly support kernels whose data dependence resolves inside the kernel \textit{and} that run for less than one HBF latency (20\,$\mu$s). Such tensors (e.g., data-dependent gathers) remain in HBM. Kernels with arbitrary tile-to-CTA mappings may also elude the profiler and validator, and require the staging warp to emulate the accesses explicitly.}

\subsection{\revhl{End-to-end Serving Throughput}}
\label{sec:serving}

Serving engines shard large models across GPUs. With each rank staging its own shard, \papername retains 87.0--104.3\% of ideal-HBF on Llama-405B at 2--8 GPUs with a 40\,MB buffer, whereas $H^3$ in address order reaches only 10.2--13.8\%. The retained fraction declines at higher degrees because kernels shorten while the 20\,$\mu$s cold start per kernel remains fixed.

\begin{figure}[htb]
\centering
\includegraphics[width=\columnwidth]{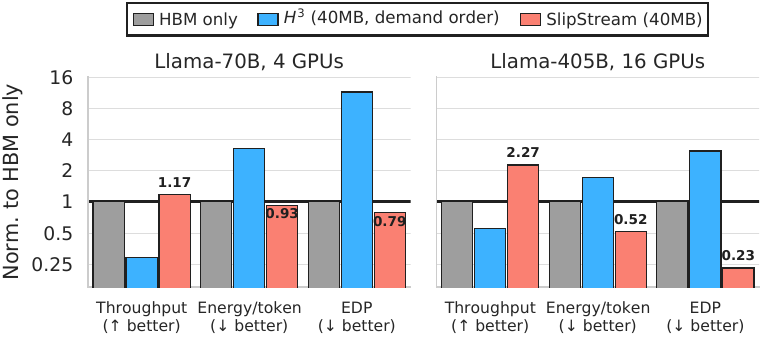}
\caption{\revhl{Serving throughput, energy per token, and EDP, normalized to HBM-only system. HBF with \papername increases lowers EDP by 1.27--4.35x.}}
\label{fig:tact_serving}
\end{figure}

\revhl{
Moving the weights to HBF frees HBM for KV cache, which allows for a larger batch. However, reading the weights from HBF can degrade time per output token (TPOT) if its latency is exposed. We derive energy per token from a 336\,W compute die, HBM at 3.97\,pJ/bit~\cite{fg_dram}, and HBF at 4$\times$ that energy~\cite{h3_llm}. For each system, we report the tensor-parallel layout with the highest max-batch throughput.
\Cref{fig:tact_serving} shows throughput, energy/token, and EDP improvements. Llama-70B fits in four HBM GPUs with sufficient KV space; \papername raises throughput by 1.17$\times$. Llama-405B needs sixteen HBM GPUs, whereas eight HBF GPUs hold a full replica. Sixteen HBF GPUs with \papername therefore deliver 2.27$\times$ the throughput, at 0.52$\times$ energy/token and 0.23$\times$ EDP. $H^3$, even in demand order, serves Llama-405B at a TPOT of 602\,ms, 10$\times$ that of HBM-only serving, and just 0.55$\times$ the throughput. Thus, HBF can improve serving throughput and energy efficiency when prefetcher hides HBF latency.
}

\section{Related Work}
\label{sec:related_work}

\subsection{GPU prefetching}
\label{sec:prefetch_related}

Prior GPU prefetchers~\cite{lee_mta,koo_cta_aware,ganguly_spatial,snake} target HBM-scale latency and track each prefetched line, so a 20\,$\mu$s lookahead exhausts their tracking state. \revhl{CABA runs assist warps on idle SM resources to relieve bottlenecks such as memory bandwidth~\cite{caba}, similar to \papername's staging warp. At HBF latency, however, a mispredicted or mistimed page stalls an SM for microseconds. The staging warp must therefore replay exact page streams under page-granular flow control.}

\begin{figure}[htb]
\centering
\includegraphics[width=\columnwidth]{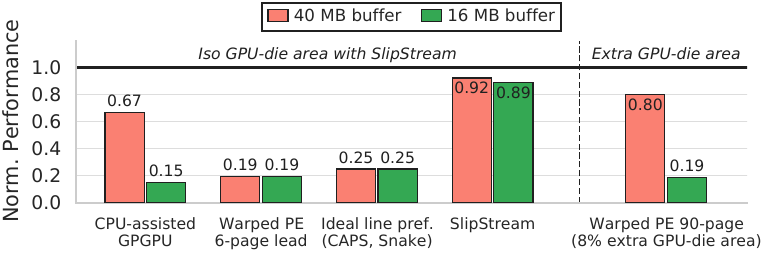}
\caption{\revhl{Prior prefetchers adapted to HBF, normalized to ideal-HBF. All designs that fit within the GPU-die area of \papername but incur 1.5-5.2x slowdown, while Warped PE with 90-page lead requires 8\% extra die area.}}
\label{fig:hbf_prefetchers}
\end{figure}

\revhl{
\Cref{fig:hbf_prefetchers} compares four prior works for HBF within the GPU-die area of \papername (geomean over seven models). Each design fetches 4\,KB pages into the staging buffer, replays exact per-SM page streams of \papername-SW, and keeps its in-flight requests and flow control. The designs fail for two reasons.

First, three designs keep too few requests in flight. Warped Pre-execution (Warped PE)~\cite{warped_pe} runs warps ahead on spare registers, which bound its lead to about \checkvalue{6} pages per SM, far short of the 90 pages that hide 20\,$\mu$s. It reaches 19.3\% of ideal-HBF. CAPS~\cite{koo_cta_aware} and Snake~\cite{snake} track every prefetched line, so a 256-entry budget, equal to that of an SSHR, covers only 8 pages per SM, so they incur 4x slowdown.

Second, designs with enough lead lack demand pacing. CPU-assisted GPGPU~\cite{cpu_gpgpu} precomputes the page streams on the CPU and issues them as one global stream. It reaches only 66.8\% at 40\,MB and 14.9\% at 16\,MB, because pages for lagging SMs hold buffer slots while other SMs starve. A 90-page lead for Warped PE needs twice the L1 capacity and register file (8\% more die area). This lead reaches 80.3\% at 40\,MB but only 18.8\% at 16\,MB, because without pacing it overruns the smaller buffer.

Finally, for \papername, using page-granular tracking alone is insufficient. Releasing each page entry on fill (like MSHRs) rather than at first demand lets the stream outpace consumption and evict pages before use. For example, on Llama-405B, 13\% of pages then require refetching, and performance drops by 9\%. Thus, our key insight of using first-demand release is essential to maximize performance (89--92\% of ideal-HBF), 
}

\subsection{Memory Expansion Techniques}
\label{sec:offload_related}

\revhl{Several systems move tensors among GPU, CPU, and storage~\cite{sheng2023flexgen,alizadeh2024llm,fusco2024understanding,moe_offloading,xue2024moeinfinity}, prefetch experts over CPU--GPU links~\cite{cao2025moe-lightning,yuzuguler2025preserve,hwang2024pregated}, or prefetch tensors or pages into HBM from host memory or flash~\cite{vdnn,capuchin,g10,deepum}. These techniques stage each tensor in HBM before its kernel starts (tensor fits in HBM). \papername instead stages its pages in a 16--40\,MB buffer that is orders of magnitude smaller than HBM. Such a buffer requires exact access order and fine-grained flow control to avoid misses.}

\subsection{Data Orchestration}
\label{sec:tma_related}

\revhl{CUTLASS and TMA pipelines stream tiles through shared memory in demand order~\cite{tiling_cutlass,cudadma}, and Buffets provide decoupled staging~\cite{pellauer2019buffets}. Both reserve buffer storage for each request at issue and release it after consumption, so storage grows with the bandwidth--delay product, which is impractical at 20\,$\mu$s. \papername instead tracks in-flight pages with compact SSHR entries and allocates storage only when a page arrives.} Staging is a non-binding hint, first demand paces admission, and final demand reclaims storage.

\section{Conclusion}
\label{sec:conclusion}

High-Bandwidth Flash (HBF) offers higher capacity at HBM-like bandwidth, but its 20\,$\mu$s read latency requires 20$\times$ more outstanding data than current GPUs support. \papername, a software--hardware co-designed prefetcher, replays the profiled page-demand order of LLM kernels ahead of demand. Its hardware retains each page-granular tracking entry until first demand, so demand paces staging, and allocates buffer storage only when pages arrive. With a 40\,MB buffer, \papername achieves 92.4\% of ideal-HBF on seven dense models (88.8\% with 16\,MB) and \revhl{91.2\% on two MoE models}. In contrast, $H^3$ incurs a 2.3$\times$ slowdown even with a 320\,MB buffer. \papername thus makes HBF practical for GPUs.

\section*{Acknowledgments}

We thank the anonymous reviewers of HPCA-2027 for their valuable feedback
We thank Dr. Po-An Tsai (NVIDIA Research) and Prof. Moinuddin Qureshi (Georgia Tech)
for insightful discussions.
Anish was partly funded by the NVIDIA Graduate Fellowship.

\bibliographystyle{IEEEtran}
\bibliography{refs}

\section*{AI Use}
The authors have used generative AI tools, Claude and ChatGPT,
in this paper to improve readability, reduce text length,
and format figures and tables.

\end{document}